\documentclass[12pt,a4paper]{article}
\usepackage{amsfonts}
\usepackage{amssymb}
\usepackage{amsmath}

\usepackage{epsfig}
\usepackage{epstopdf}
\usepackage{graphicx}

\usepackage[superscript]{cite}
\usepackage[breaklinks]{hyperref}  % for pdfLaTeX users
\hypersetup{colorlinks,urlcolor=black,citecolor=black,linkcolor=black,filecolor=black}
\usepackage{breakurl}

\date{}

\begin{document}

\author{\small{Ya-Bo Wu\footnote{\href{mailto:ybwu61@163.com} {Corresponding author: ybwu61@163.com}},
Xue Zhang, Bo-Hai Chen, Nan Zhang and Meng-Meng Wu}}

\title{Energy conditions and constraints on the generalized non-local gravity model}
\maketitle

{\small{\centerline{\small{Department of Physics,
Liaoning Normal University, Dalian 116029, P.R.China}}

\begin{abstract}
We study and derive the energy conditions in generalized non-local gravity,
which is the modified theory of general relativity (GR) obtained
by adding a term $m^{2n-2}R\Box^{-n}R$ to the Einstein-Hilbert action.
Moreover, in order to get some insight on the meaning of the energy conditions,
we illustrate the evolutions of four energy conditions with the model parameter $\varepsilon$ for different $n$.
By analysis we give the constraints on the model parameters $\varepsilon$.



\end{abstract}

\section{Introduction}
\label{sec:1}

According to recent observational data sets \cite{astro-ph/9812133,astro-ph/9805201,arXiv:1212.5226,arXiv:1502.01589},
it has been believed that our universe is flat and undergoing a phase of the accelerating expansion at the present time.
Explaining this problem is a challenge of modern cosmology.
As we know the Einstein equation govern the interplay between the geometry of the space-time and the matter distribution.
Different ways addressing the problem of accelerating expansion can be classified as the gravity side or the matter side of the Einstein equation.
On the one hand, it is reasonable to believe that the acceleration of the Universe is probably driven by dark energy,
which is an exotic component with negative pressure and provides the main contribution to the energy budget of the universe today.
The simplest candidate for the dark energy is the cosmological constant $\Lambda$ \cite{astro-ph/9904398,astro-ph/0004075},
but the cosmological observations indicate that the dark energy may not be a constant.
Several candidates for dark energy have been extensively investigated \cite{astro-ph/0207347,hep-th/0603057,arXiv:0803.0982,arXiv:1103.5870}.
Unfortunately, a satisfactory answer to the questions of what dark energy is and where it comes from has not yet been obtained.
On the other hand, as an alternative to dark energy, modified theories of gravity are extremely attractive
(see Refs.~11-14 %\cite{arXiv:1011.0544,arXiv:1106.2476,arXiv:1108.6266,U2}
for reviews).
The challenge is to theoretically construct a consistent theory more effectively explaining the acceleration data,
and the significant deviations from GR are neglected by the data inside the solar system.
There are numerous ways to modify the Einstein's theory of GR.
Among these theories, $f(R)$ gravity is one of the competitive candidates in modified theories of gravity \cite{arXiv:0805.1726,arXiv:1002.4928},
in which $f(R)$ is an arbitrary function of the Ricci scalar $R$.
Another interesting alternative modified theory of gravity is the non-local gravity model proposed by Deser and Woodard \cite{arXiv:0706.2151}.
The model is constructed by adding a non-local term of $f(\Box^{-1}R)$ to the Einstein-Hilbert action.
$\Box^{-1}$ operator is a formal inverse of the d'Alembertian $\Box$ in the scalar representation
which can be expressed as the convolution with a retarded Green's function \cite{arXiv:1307.3898,arXiv:1112.4340,arXiv:1107.1463,arXiv:0910.4097,arXiv:0809.4927}.
The form of the non-local distortion function $f(\Box^{-1}R)$ can be chosen to reproduce the $\Lambda$CDM background cosmology exactly \cite{arXiv:0803.3399,arXiv:0904.0961}.
The Ricci scalar $R$ vanishes during radiation dominance
and $\Box^{-1}R$ cannot begin to grow until the onset of matter dominance
while its growth becomes logarithmic.
Moreover, the great advantage of this class of models is to trigger late time acceleration by the transition from radiation domination.
This theory have attracted some attention both theoretically and phenomenologically,
as a possible alternative to dark energy that presents the universe accelerated at late times.
The model can be compared with observations, but fails miserably to account for structure formation data.
Furthermore, another new non-local model has been proposed by Maggiore and Mancarella (M-M),
which is by adding a specific form $m^2\Box^{-2}R$ to the Einstein-Hillbert action \cite{arXiv:1402.0448}.
A natural way to proceed is to introduce a mass scale $m$ which is in order of the Hubble parameter' present value $H_0$.
In contrast with the Deser-Woodard model,
the non-local term is controlled by a mass parameter $m$.
Currently this model is receiving more attention \cite{arXiv:1408.5058,arXiv:1403.6068,arXiv:1408.1084},
because it can be compared with a wide set of observations
\cite{arXiv:0807.3778,arXiv:1110.5249,arXiv:1209.0836,arXiv:1305.3034,arXiv:1311.3435,arXiv:1307.6968,arXiv:1402.4613}.
and it is so far the only model which is as good as $\Lambda$CDM, using the same number of free parameters.
Recently, we extend the model to the generalized non-local (GNL) gravity, which is defined by the action \cite{arXiv:1511.05238}
\begin{eqnarray}\label{GNL}
S_\mathrm{GNL}=\frac{1}{16\pi G}\int d^{d+1}x\sqrt{-g}R \Big(1-\lambda\frac{m^{2n-2}}{\Box^{n}}R \Big),
\end{eqnarray}
where $n$ takes integer ($n\geq2$), $\lambda=(d-1)/4d$ is the normalized coefficient and $d$ is the number of spatial dimensions.
the convenient normalization of the mass parameter $m$ which is in order of the Hubble parameter present value $H_0$.
A natural way to proceed is to set the mass scale $m=\varepsilon H_0$, where $\varepsilon$ is a constant, which is here called as the model parameter.
Thus, when considering four-dimensional spacetime (i.e. $d=3$), the action \eqref{GNL} can be rewritten as
\begin{eqnarray}\label{GNLaction}
S_\mathrm{GNL}=\frac{1}{16\pi G}\int d^{4}x\sqrt{-g}R \Big(1-\frac{H_0^{2n-2}}{6}\varepsilon^{2n-2}\Box^{-n}R \Big).
\end{eqnarray}

On the other hand, since many models of modified theories of gravity have been proposed,
it gives rise to the problem how to constrain them from theoretical aspects.
One possibility is by imposing the so-called energy conditions \cite{book,gr-qc/9510008,gr-qc/0608072,arXiv:0708.0411,arXiv:1011.4159,arXiv:1111.3878,arXiv:1207.1503,arXiv:1211.3740,arXiv:1212.4656,arXiv:1302.0466,arXiv:1306.3450}, namely,
the strong energy condition (SEC),
the null energy condition (NEC),
the dominant energy condition (DEC)
and the weak energy condition (WEC).
In this letter, we tackle the problem of the energy conditions in generalized non-local gravity.
Considering these energy conditions, one is allowed not only to establish gravity which remains attractive,
but also to keep the demands that the energy density is positive and cannot flow faster than light.
This issue is extremely delicate since a standard approach is to consider the gravitational field equations as effective Einstein equations.
In order to obtain the SEC and NEC, the Raychaudhuri equation which is their physical origin is used.
It is worth stressing that the equivalent results can be obtained by taking the transformations
$\rho\rightarrow\rho^{\text{eff}}$ and $p\rightarrow p^{\text{eff}}$ into $\rho+3p\geqslant0$ and $\rho+p\geqslant0$, respectively.
Thus by extending this approach to $\rho-p\geqslant0$ and $\rho\geqslant0$, the DEC and WEC are obtained.
As is well known, at present the energy conditions have been used in various modified gravity widely
\cite{arXiv:0903.4540,arXiv:1212.4921,arXiv:1212.4928,arXiv:1203.5593,yyzMPLA}.
Thus, we wonder if the energy conditions for the generalized non-local
gravity model as well as the constraints on the model could be
obtained, which is our motivation in this letter. The research
results show that the energy conditions in generalized non-local
gravity can be obtained which can degenerate to the ones in GR as
special cases. Moreover, the constraints on the model parameter
$\varepsilon$ can be given by means of the SEC, the NEC and DEC.

\section{Field equations of generalized non-local gravity}
\label{sec:2}

We consider the generalized non-local gravity model given by the action \eqref{GNLaction},
where the covariant scalar d'Alembertian is
\begin{equation}
\Box\equiv\frac{1}{\sqrt{-g}}\partial_{\mu}\big(\sqrt{-g}g^{\mu\nu}\partial_{\nu}\big).
\end{equation}
In general, the nonlocal term was solved by the retarded Green's function via \cite{arXiv:1307.3898,arXiv:1408.5058}
\begin{equation}
(\Box^{-1}R)(x)\equiv\int d^4x\sqrt{-g(x')}G(x, x')R(x'),
\end{equation}
where $G(x, x')$ is the Green's function of the $\Box^{-1}$ operator acting on scalar.
We consider that the non-local effect only starts at cosmic time $t_0$, so we must impose the boundary conditions.
For such a $G(x, x')$ we choose initial time to lie in deep radiation dominated period.
Indeed, during radiation dominated period $R=0$, which accords with the condition in general relativity.
Incidentally, either non-local terms $R\Box^{-2}R$ or $(\Box^{-1}R)^2$ in the action will give the same equations of motion
and this effectively translates in the freedom of integrating by parts $\Box^{-1}$.

Taking the variation of the action \eqref{GNLaction} with respect to the metric $g^{\mu\nu}$,
we obtain the modified Einstein equations
\begin{equation}\label{Einstein1}
G_{\mu \nu}-\frac{H_0^{2n-2}}{6} \varepsilon^{2n-2}K_{\mu \nu}=8\pi GT_{\mu \nu}
\end{equation}
with
\begin{align}\nonumber
K_{\mu \nu}=&\Big[2(G_{\mu \nu}-\nabla_{\mu}\nabla_{\nu}+g_{\mu \nu}\Box)+\frac{1}{2}g_{\mu \nu}R\Big](\Box^{-n}R)\\\nonumber
&+\sum^{n-1}_{l=0}\bigg\{\nabla_{\nu}(\Box^{l-n}R)\nabla_{\mu}(\Box^{-l-1}R)\\\label{Einstein2}
&-\frac{1}{2}g_{\mu \nu}\Big[\nabla_{\sigma}(\Box^{l-n}R)\nabla^{\sigma}(\Box^{-l-1}R)+(\Box^{l-n}R)(\Box^{-l}R)\Big]\bigg\},
\end{align}
where $\Box=g^{\mu\nu}\nabla_{\mu}\nabla_{\mu}$, $K_{\mu \nu}$ comes from varying the non-local term in the above action
and the energy-momentum tensor of matter $T_{\mu \nu}$ is defined by
\begin{equation}
T_{\mu\nu}=-\frac{2}{\sqrt{-g}}\frac{\delta(\sqrt{-g}\mathcal{L}_\text{M})}{\delta g^{\mu\nu}}.
\end{equation}

We need to rewrite the original GNLG model as a local form,
Following the same definitions and localization procedure in Ref. 25 %\cite{arXiv:1402.0448}
we introduce the following auxiliary scalar fields $U_1, \cdots, U_n$
\begin{align}\label{10}
U_{1}=&-\Box^{-1}R,\\
U_{2}=&-\Box^{-1}U_{1}=\Box^{-2}R,\\
U_{3}=&-\Box^{-1}U_{2}=-\Box^{-3}R,\\
U_{4}=&-\Box^{-1}U_{3}=\Box^{-4}R,\\\nonumber
&\vdots\\\label{14}
U_{n}=&-\Box^{-1}U_{n-1}=(-1)^n\Box^{-n}R,
\end{align}
whose initial conditions are all zero.
For convenience, we set a set of variables $V_{i}= H_{0}^{2}U_{i}~(i=2, \cdots, n)$.
One reads a set of coupled differential equations
\begin{align}\label{15}
U_1''+U_1'(3+\zeta)&=6(2+\zeta),\\
V_2''+V_2'(3+\zeta)&=h^{-2}U_1,\\
V_3''+V_3'(3+\zeta)&=h^{-2}U_2,\\
V_4''+V_4'(3+\zeta)&=h^{-2}U_3,\\\nonumber
\vdots&\\\label{19}
V_n''+V_n'(3+\zeta)&=h^{-2}U_{n-1}.
\end{align}
Here $h=H(t)/H_{0}$,
where $H(t)=\dot{a}/a$ and $H_{0}$ is the present value of the Hubble parameter. $\zeta=h'/h$
and a prime denotes the derivative with respect to the time coordinate $\ln{a}$.

We consider a flat FRW background with metric
\begin{equation}
ds^2=-dt^2+a^2(t)d\mathbf{x}^2.
\end{equation}
From the (00) component of the modified Einstein equation \eqref{Einstein1}, we obtain
\begin{equation}
K_{00}=(-1)^n\Big[6h^{2}V'_{n}+6h^{2}V_{n}-\frac{1}{2}h^{2}\Theta_1+\frac{1}{2}\Theta_2\Big],
\end{equation}
where
\begin{align}
\Theta_1&=\sum^{n-1}_{l=0}V_{n-l}'U_{l+1}',\\
\Theta_2&=\sum^{n-1}_{l=1}U_{n-l}U_l,
\end{align}
as well as
\begin{align}
\Theta_1'&=\sum^{n-1}_{l=0}\Big[-2(3+\zeta)V_{n-l}'U_{l+1}'-h^{-2}(U_{n-l-1}U_{l+1}'+U_{n-l}'U_{l})\Big],\\
\Theta_2'&=\sum^{n-1}_{l=1}(U_{n-l}'U_l+U_{n-l}U_l').
\end{align}
Considering these definitions, one reads an effective dark energy
density $\rho_{DE}=\rho_0\beta Y$ where $\rho_0 =3H^2_0/(8\pi G)$,
$\beta=\frac{H_0^{2n-4}}{9}\varepsilon^{2n-2}$ and
\begin{equation}
Y=(-1)^n \Big(3h^{2}V'_{n}+3h^{2}V_{n}-\frac{1}{4}h^{2}\Theta_{1}+\frac{1}{4}\Theta_{2}\Big).
\end{equation}
Thus, we get
\begin{equation}\label{h2}
h^{2}=\frac{\Omega_M e^{-3x}+\Omega_R e^{-4x}+\frac{1}{4}\beta(-1)^n \Theta_2}{1-\beta(-1)^n(3V_n'+3V_n-\frac{1}{4}\Theta_1)}
\end{equation}
and
\begin{equation}
\zeta=\frac{h^{-2}(-3\Omega_M e^{-3x}-4\Omega_R e^{-4x})-3\beta(-1)^n(-h^{-2}U_{n-1}+4V_n'-\frac{1}{2}\Theta_1)}{2\big[1-3\beta(-1)^n V_n\big]},
\end{equation}
where $\Omega_M$, $\Omega_R$ are the fractional energy densities of matter and radiation, respectively.
Also, with $\zeta$ and $h^2$, Eqs.~\eqref{15} and \eqref{19} provide a closed set of second-order differential equations for $V_{i}= H_{0}^{2}U_{i}~(i=2, \cdots, n)$ ,
whose numerical integration is straightforward.

\section{Energy conditions and constraints on the generalized non-local gravity model}
\label{sec:3}

\subsection{Energy conditions}

We consider the energy conditions in the generalized non-local gravity.
When $n$ is taken as arbitrary interger ($n\geq2$), Eq. \eqref{Einstein1} may be rewritten as the following effective gravitational field equation
\begin{equation}\label{16}
G_{\mu\nu}\equiv R_{\mu\nu}-\frac{1}{2}g_{\mu\nu}R=T_{\mu\nu}^\text{eff},
\end{equation}
where
\begin{align}\nonumber
T_{\mu \nu}^\text{eff}=T_{\mu \nu}&+(-1)^n\frac{H_0^{2n-2}}{48\pi G} \varepsilon^{2n-2}
\bigg\{\Big[2(G_{\mu \nu}-\nabla_{\mu}\nabla_{\nu}+g_{\mu \nu}\Box)+\frac{1}{2}g_{\mu \nu}R\Big]U_n\\
&-\sum^{n-1}_{l=0}\Big[\nabla_{\nu}U_{n-l}\nabla_{\mu}U_{l+1}
-\frac{1}{2}g_{\mu \nu}(\nabla_{\sigma}U_{n-l}\nabla^{\sigma}U_{l+1}-U_{n-l}U_{l})\Big]\bigg\}.
\end{align}
Moreover,
\begin{align}
\rho^\text{eff}=&-T_{00}^\text{eff}g^{00}
=\rho+(-1)^n\frac{H_0^{2n-4}\rho_{0}}{18} \varepsilon^{2n-2}
(6h^2V_n+6h^2V'_n-\frac{1}{2}h^2\Theta_1+\frac{1}{2}\Theta_2),\\
p^{\text{eff}}=&\frac{1}{3}T_{ii}^\text{eff}g^{ii}
=p+(-1)^n\frac{H_0^{2n-4}\rho_{0}}{18} \varepsilon^{2n-2}
(-4h^2 \zeta V_n-6h^2 V_n+2h^2 V'_n-2U_{n-1}-\frac{1}{2}h^2 \Theta_1-\frac{1}{2}\Theta_2),
\end{align}
where energy density $\rho=\rho_{\text{M}}+\rho_{\text{R}}$ and $p=p_{\text{R}}$ and $\rho_0 =3H^2_0/(8\pi G)$.

Note that the SEC and NEC are directly derived from Raychaudhuri equation
and the equivalent expressions can be obtained by taking the transformations
$\rho\rightarrow\rho^{\text{eff}}$ and $p\rightarrow p^{\text{eff}}$ into $\rho+3p\geqslant0$ and $\rho+p\geqslant0$, respectively.
\cite{arXiv:0708.0411,arXiv:1011.4159,arXiv:1111.3878,arXiv:1207.1503,arXiv:1211.3740,arXiv:1212.4656,
arXiv:1302.0466,arXiv:1306.3450,arXiv:0903.4540,arXiv:1212.4921,arXiv:1212.4928,arXiv:1203.5593,yyzMPLA}.
By extending this approach to $\rho-p\geqslant0$ and $\rho\geqslant0$,
we can give the corresponding DEC and WEC in the generalized non-local gravity.
Hence, the four energy conditions, i.e., SEC, NEC, DEC, WEC,
in generalized non-local gravity can be respectively given as
\begin{align}\nonumber
\text{SEC}:~&\rho^{\text{eff}}+3p^{\text{eff}}\geq0\Rightarrow\\\nonumber
&\Omega_\text{M}+\Omega_\text{R}(1+3w_\text{R})+(-1)^n\frac{H_0^{2n-4}}{18}\varepsilon^{2n-2}
(-12h^2\zeta V_n-12h^2V_n+12h^2V'_n-6U_{n-1}\\\label{SEC2468}
&-2h^2\Theta_1-\Theta_2)\geq0,\\\nonumber
\text{NEC}:~&\rho^{\text{eff}}+p^{\text{eff}}\geq0\Rightarrow\\\label{NEC2468}
&\Omega_\text{M}+\Omega_\text{R}(1+w_\text{R})+(-1)^n\frac{H_0^{2n-4}}{18}\varepsilon^{2n-2}
(-4h^2\zeta V_n+8h^2V'_n-2U_{n-1}-h^2\Theta_1)\geq0,\\\nonumber
\text{DEC}:~&\rho^{\text{eff}}-p^{\text{eff}}\geq0\Rightarrow\\\label{DEC2468}
&\Omega_\text{M}+\Omega_\text{R}(1-w_\text{R})+(-1)^n\frac{H_0^{2n-4}}{18}\varepsilon^{2n-2}
(4h^2\zeta V_n+12h^2V_n+4h^2V'_n+2U_{n-1}+\Theta_2)\geq0,\\\nonumber
\text{WEC}:~&\rho^{\text{eff}}\geq0\Rightarrow\\\label{WEC2468}
&\Omega_\text{M}+\Omega_\text{R}+(-1)^n\frac{H_0^{2n-4}}{18}\varepsilon^{2n-2}
(6h^2V_n+6h^2V'_n-\frac{1}{2}h^2\Theta_1+\frac{1}{2}\Theta_2)\geq0.
\end{align}

\subsection{Constraints on the model parameter $\varepsilon$}

In order to deeply understand the above inequalities, we have
plotted the evolutions of the four energy conditions, i.e., Eqs.
\eqref{SEC2468}-\eqref{WEC2468} with the model parameter
$\varepsilon$ for different $n$ in Fig. \ref{fig:1}.
For simplicity, we chose to use $n=2,\cdots,8$.
Here we take $\Omega_\text{M}=0.308$, $\Omega_\text{R}=0.001$ and $w_\text{R}=1/3$ \cite{arXiv:1502.01589}.
From Fig. \ref{fig:1} it is not difficult to see that for the different even number $n$ the SEC and the NEC
can give the constraints on the model parameters $\varepsilon$, respectively.
Whereas for the different odd number $n$ the DEC give the constraints on the model parameters $\varepsilon$.
In addition, Table \ref{tab:1} shows the constraints on the model
parameters $\varepsilon$ for different numbers of $n$. Here the
symbol "-" stands for no limits to the model parameter
$\varepsilon$. From Table \ref{tab:1} we can easily see that when
taking the even numbers of $n$, the bigger $n$ is, the wider value
range of $\varepsilon$ is, which is determined by the SEC and the
NEC, while the DEC and WEC can not give any constraints on the
model parameter $\varepsilon$. In the other words, the DEC and WEC
are always satisfied in any value ranges of the model parameter
$\varepsilon$. However, when taking the odd numbers of $n$, the
DEC constrains almost the same value ranges on $\varepsilon$ in
this model, while other three energy conditions (i.e., SEC, NEC
and WEC) can not give any constraints on the model parameter
$\varepsilon$.

\begin{figure}[!htbp]
  \centering
  \includegraphics[width=.6\textwidth]{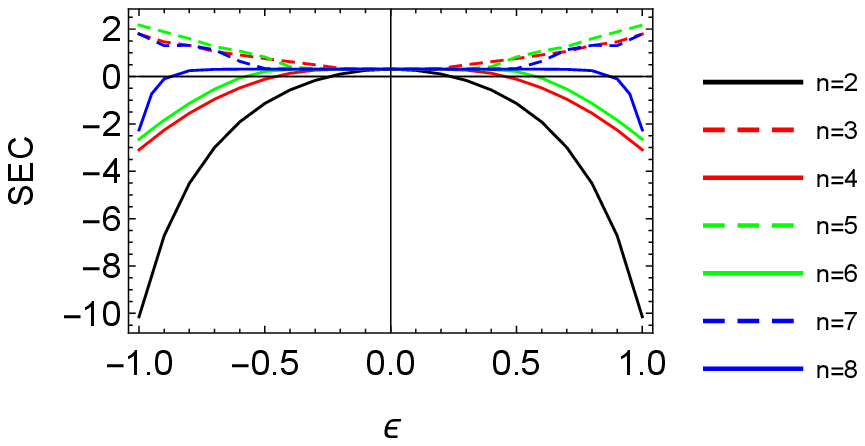}
  \includegraphics[width=.6\textwidth]{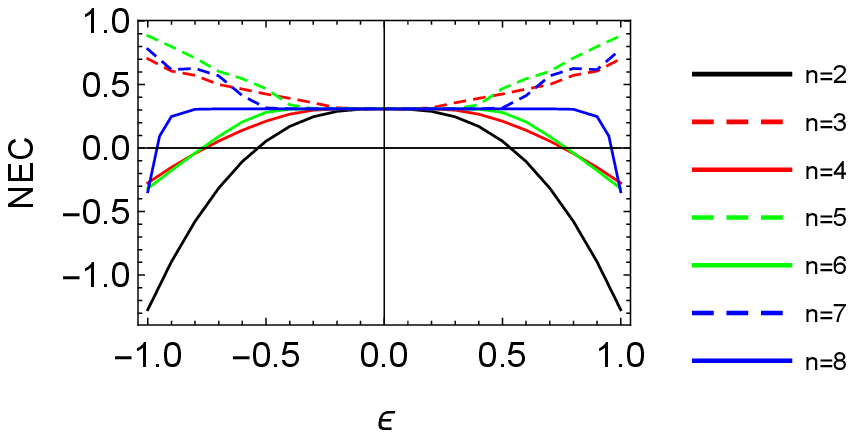}
  \includegraphics[width=.6\textwidth]{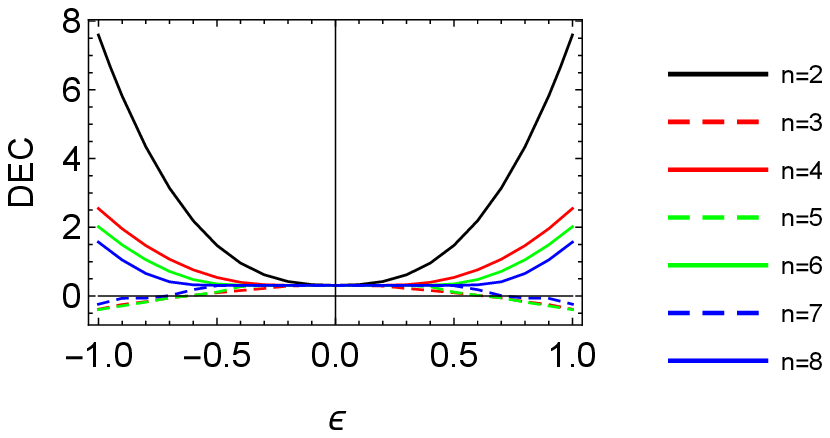}
  \includegraphics[width=.6\textwidth]{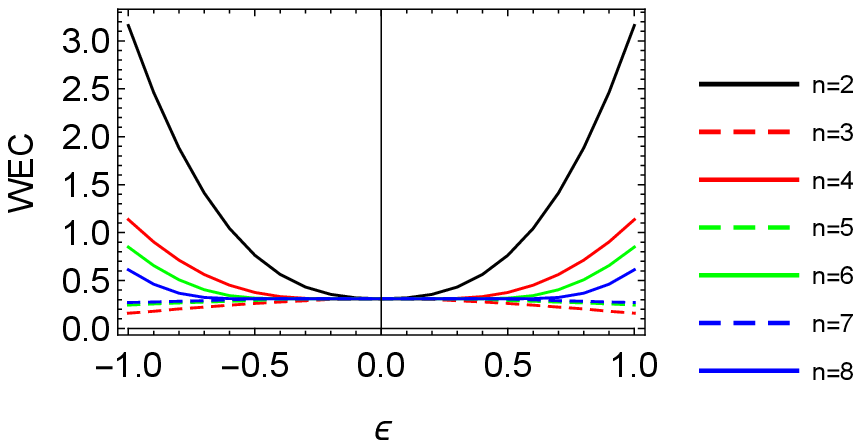}
  \caption{The evolutions of SEC, NEC, DEC, WEC with the model parameter $\varepsilon$ for different $n~(n=2,\cdots,8)$.}\label{fig:1}
\end{figure}

\begin{table}[!htbp]
\centering
\begin{tabular}{|c|c|c|c|c|c|}
  \hline
        & SEC                & NEC                & DEC & WEC & SEC\&NEC\&DEC\&WEC\\
  \hline
  $n=2$ & $|\varepsilon|\leq0.26$ & $|\varepsilon|\leq0.53$ & -   & -   & $|\varepsilon|\leq0.26$ \\
  $n=3$ & - & - & $|\varepsilon|\leq0.63$ & -   & $|\varepsilon|\leq0.63$ \\
  $n=4$ & $|\varepsilon|\leq0.46$ & $|\varepsilon|\leq0.75$ & -   & -   & $|\varepsilon|\leq0.46$ \\
  $n=5$ & - & - & $|\varepsilon|\leq0.62$ & -   & $|\varepsilon|\leq0.62$ \\
  $n=6$ & $|\varepsilon|\leq0.58$ & $|\varepsilon|\leq0.77$ & -   & -   & $|\varepsilon|\leq0.58$ \\
  $n=7$ & - & - & $|\varepsilon|\leq0.72$ & -   & $|\varepsilon|\leq0.72$ \\
  $n=8$ & $|\varepsilon|\leq0.88$ & $|\varepsilon|\leq0.96$ & -   & -   & $|\varepsilon|\leq0.88$ \\
  \hline
\end{tabular}
\caption{\label{tab:1}  The constraints on the model parameter $\varepsilon$ by the four energy conditions for different $n~(n=2,\cdots,8)$.}
\end{table}

\section{Summary}
\label{sec:4}

So far, we have derived the well known strong energy condition, the null energy condition, the dominant energy condition
and the weak energy condition for the generalized non-local gravity model,
which is obtained by adding a term $m^{2n-2}R\Box^{-n}R$ to the Einstein-Hilbert action.
Moreover, in order to get some insight on the meaning of the energy conditions,
we have illustrated the evolutions of the four energy conditions
in terms of the model parameter $\varepsilon$ for different $n$ in Fig.\ref{fig:1},
and as shown in table \ref{tab:1} we have given the constraints on the model parameters $\varepsilon$ for different $n$
in generalized non-local gravity model satisfying the SEC, NEC, DEC and WEC, respectively.
From Fig. \ref{fig:1} it is not difficult to see that for the different even number $n$ the SEC and the NEC
can give the constraints on the model parameters $\varepsilon$, respectively.
Whereas for the different odd number $n$ the DEC give the constraints on the model parameters $\varepsilon$.
From Table \ref{tab:1} we can easily see that
when taking the even numbers of $n$,
the bigger $n$ is, the wider value range of $\varepsilon$ is,
which is determined by the SEC and the NEC,
while the DEC and WEC can not give any constraints on the model parameter $\varepsilon$.
However, when taking the odd numbers of $n$, the DEC constrains almost the same value ranges on $\varepsilon$ in this model, while
SEC, NEC and WEC can not give any constraints on the model parameter $\varepsilon$.

\section*{Acknowledgments}

This work is supported by the National Natural Science Foundation of China (Grant Nos. 11175077 and 11575075)
and the Natural Science Foundation of Liaoning Province in China (Grant No. L201683666).

\end{document}